\documentclass[preprint,amsmath,amssymb]{revtex4}
\usepackage{latexsym}
\usepackage[dvips]{graphicx}
\usepackage{bm}

\newcommand{\beq}{\begin{equation}}
\newcommand{\eeq}{\end{equation}}
\newcommand{\ba}{\begin{array}}
\newcommand{\ea}{\end{array}}
\newcommand{\beqa}{\begin{eqnarray}}
\newcommand{\eeqa}{\end{eqnarray}}
\begin{document}

\title{Bilepton and exotic quark mass limits in 331 models from $Z \rightarrow b\bar{b}$ decay}
\author{G. A. Gonz\'alez-Sprinberg (1), R. Mart\'\i nez (2) and  O. Sampayo (3)}

\altaffiliation{gabrielg@fisica.edu.uy,
remartinezm@unal.edu.co, sampayo@mdp.edu.ar}
\affiliation{(1) Instituto de F\'{\i}sica,
 Facultad de Ciencias, Universidad de la Rep\'ublica,  Igu\'a 4225,\\
11400 Montevideo, Uruguay\\ \\
(2) Departamento de F\'\i sica, Universidad Nacional de Colombia \\
Bogot\'a, Colombia\\ \\
(3) Departamento de F\'\i sica, Universidad Nacional de Mar del Plata,\\
Funes 3350, (7600) Mar del Plata, Argentina}

\date{February, 2005}


\begin{abstract}
We study  the effect of new physics on the Z-decay into $b\bar{b}$
pairs in the framework of 331 models. The decay $Z \to b \bar{b}$
is computed at one loop level and,  using previous results, we
evaluate this branching fraction in the framework of  331 models.
A wide range of the space parameter of the model is considered and
possible deviations from the standard model predictions are
explored. From precision measurements at the  $Z$-pole we find the
allowed region for $M_{J_3}$, $M_\chi$ at $95\%$ CL.

\end{abstract}

\maketitle

\section{Introduction}
The Standard Model (SM) \cite{sm}
has passed almost all the experimental test in the past,
except for the recently discovered neutrino oscillation \cite{no}.
The SM predictions have been extensively checked, for example,  in
 the neutral currents phenomenology, the
 GIM mechanism \cite{gim} and the invisible decay of the
Z, where three light neutrino families can be accommodated with
experimental data.
 One loop quantum corrections at the $Zb\bar{b}$ vertex \cite{zbb} have shown the
first hint of the top-quark mass, latter discovered at FERMILAB \cite{top}.
Moreover, oblique corrections show that the Higgs´s mass value should be
at the level of the electroweak scale \cite{higgs}.

However, there are some  fundamental questions  that remain
unanswered within the SM framework. The first experimental clue
came from neutrino oscillations as an evidence for some physics
beyond the SM, but the SM also has no answer to the fermion and
neutrino mass scales and mixing angles, the number of families,
the origin of electric charge quantization, among a list of
important
 questions \cite{preg}. Some of these problems may be a hint to new physics
 above  the electroweak scale and below
the Planck scale and are addressed in extended models using
different theoretical ideas. For example, the naturalness problem
is solved
 in supersymmetric theories. Quadratic quantum corrections to the Higgs mass
can be rendered stable in supersymmetric extensions of the SM, in
particular the Minimal Supersymmetric SM (MSSM) where the Higgs
mass can be light\cite{higgslight}. The stability of the Higgs
mass can also be obtained in the recently proposed Little Higgs
theories \cite{lh}. On the other hand, Warped Extra Dimension
theories can explain the fine tuning between the electroweak scale
and the Planck scale \cite{randall}. Within these frameworks one
can naturally
 have symmetry breaking without
scalar fields.

The so-called (331) models, based on the $SU(3)_C\otimes
SU(3)_L\otimes U(1)_X$ gauge group, has been recently proposed
\cite{m331}. In these models the cancellation of anomalies
 imposes to have three fermion
families. There are some models with one and  three families where
this number is regulated
 by the value of the parameter $\beta$ in the definition
of the electromagnetic charge\cite{flias}. In three families
models, there are two classes of them studied in the literature
with $\beta=-\sqrt{3}$, $\beta=1/\sqrt{3}$, where the first of
them has bileptons in the spectrum. These models can be tested in
the near future at the LHC and
 have an interesting phenomenology. Besides bileptons they
also predict the existence of  exotic quarks. There are some
studies on the mass limits for these particles
\cite{d1,d2,d3,d4,d5}. The exchange of doubly charged  bileptons
in the processes $e^{+} e^{-} \rightarrow \mu^{+}\mu^{-}$ and
$e^{+} e^{-}\rightarrow \tau^{+} \tau^{-} $ have been analyzed at
center of mass energies from $189\ GeV$ to $207\ GeV$ and it is
found that $(g_L/M_\chi)^{2}\approx 10^{-6} \ GeV^{-2}$ up to 95\%
CL \cite{d1}. From the contribution of bileptons to the oblique
parameters S and T a bound $M_\chi > 500\ GeV$ has been
established \cite{d2}; from muonium - antimuonium conversion the
$M_\chi > 850\ GeV$ bound is found \cite{d3}. Fermion pair
production in $e^{+} e^{-} $ annihilation allows to put the limit
$M_\chi > 740\ GeV$; this last bound is less stringent than the
previously mentioned but it is less dependent on free parameters
\cite{d4}. LHC production has been studied in \cite{d5}.

In this paper we want to study the allowed region for the spectrum
of the $J_3$ particles, all of them with fractional electric
charge $ \pm 5/3$, and for the bilepton $\chi^{++}$ for some (331)
models \cite{m331}. This allowed region is constrained by
comparing the data with radiative corrections to the decay $Z
\rightarrow b\bar{b}$. A $\chi^2$ study can be made taking
$\Gamma_Z$, $R_b$, $R_c$, $R_l$ and $\Gamma(Z \rightarrow b\bar{b})$ as
parameters. Both the  $A_b$ and $A_{FB}^b$ asymmetries are within
$3\sigma$ from the SM predictions and we are not going to take
these into account in our study \cite{exotic}. On the other hand,
as this model has a left-handed structure it can not resolve this
last issue at the one loop level because  the contribution to the
value of $g_L^{bbZ}$ is very small \cite{dg}. The mixing of the
neutral
 current with the additional one ({\it i.e.},the  $Z$
 and $Z'$ neutral currents) needs a mixing angle of the order of
$\sin\theta = -0.3625$ to explain the  $A_b$ and $A_{FB}^b$
deviations,
 and this value is not compatible with the value one
can obtain from the analysis of the $Z$-pole observables and the
weak charge $Q_W$, which is of the order of $\sin\theta
\le 10^{-3} - 10^{-4}$ \cite{wch}.

In the following sections we present the model, the new physics
contribution to the LEP observables and the conclusions.

\section{The 331 Model}

The fermionic states can be taken as in Ref.\cite{foot}:

\beq f_L^m = \left( \ba{c} \nu_m \\l_m \\l^+_m \ea \right)_L \sim
(1,3,0) \ ; \
 Q_3 = \left( \ba{c} t \\ b\\ J_3 \ea \right)_L \sim (3,3,2/3) \ ; \
Q^i_L = \left( \ba{c} d_i \\ -u_i \\ J_i \ea \right)_L \sim (3,
\bar{3}, -1/3) \nonumber \eeq \beq u_{a,R} \sim (3,1,5/3) \ ; \
 d_{a,R} \sim (3,1,-1/3)
\nonumber \eeq
\beq
J_{3R} \sim (3,1,5/3) \;\; ; \;\;
 J_{1,2 R} \sim (3,1, -4/3)
\eeq \noindent where $a,m = 1, 2, 3$ ; $i = 1, 2$ represent the
two light families of the SM. The exotic quarks $J_3$ and
$J_{1,2}$ have electric charge $\pm 5/3$ and $\pm 4/3$
respectively and then,  they do not mix in the mass matrix. Only
the $J_3$ quark contributes to the one loop quantum corrections to
$Z \rightarrow b\bar{b}$. This model does not have exotic leptons.
This implies that the one loop new physics contribution to $Z
\rightarrow l^-l^+$
 is suppressed with respect to the SM one. The electric charge is
\begin{equation}
Q=T_{3L}-\sqrt{3} T_{8L}+Y_N
\end{equation}
where $T_{3L}$,  $T_{8L}$ and $Y_N$ are the generators of the $SU(3)_L$ and
$U(1)_X$, respectively.

The scalar fields content, responsible for the symmetry breaking
and the fermion masses, is given by three Higgs triplets:
\begin{eqnarray}
\rho &=& \left( \ba{c} \rho^+ \\ \rho^0 \\ \rho^{++} \ea \right) \sim (1,3,1) \;\ ; \;\
\eta = \left( \ba{c} \eta^0 \\ \eta^-_1\\ \eta^+_2 \ea \right) \sim (1,3,0) \; ; \; \nonumber \\
\chi &=& \left( \ba{c} \chi^- \\ \chi^{--}\\ \chi^0 \ea \right) \sim (1,3, -1)
\end{eqnarray}
In order to give mass to the neutrino sector
one can also introduce a Higgs sextet:
\begin{equation}
S = \left( \begin{array}{ccc}
\sigma^0_1 & s_ 2^+& s_1^- \\
 s_2^+ & s_1^{++} &\sigma_2^0\\
s_1^- & \sigma_2^0 & s_2^{--}
\end{array}\right) \sim (1, \bar{6}, 0).
\end{equation}
The vacuum expectation values (VEV) for the scalar fields are:
\beq \langle\rho\rangle \sim \frac{u}{\sqrt{2}}\;\;,\;\;
\langle\eta\rangle \sim\frac{v}{\sqrt{2}}\;\;,\;\;
\langle\delta\rangle \sim \frac{v_\sigma}{\sqrt{2}}\;\;,\;\;
\langle\chi\rangle \sim \frac{w}{\sqrt{2}} \eeq
where the VEV's values are chosen to obey the following relations:
 \beq \ba{l}
w \gg u, v, v_\sigma \sim V \\
V^2 = u^2 + v^2 + v_\sigma^2 \ea .\eeq

With the mass eigenstates and the would-be Goldstone bosons coming
from the Higgs potential \cite{long, flias} one can write the
relevant couplings in order to compute the one loop correction to
$Z \rightarrow b\bar{b}$. The bosons are written as:
 \beq
\rho = \left( \ba{c} G_W^+ \\
\frac{i G_Z+V}{\sqrt{2}} \\ 0 \ea \right)  \;\ ; \;\
 \eta = \left( \ba{c} \frac{- i G_Z+V}{\sqrt{2}}
  \\  -G_W^-\\ 0 \ea \right) \; ; \;
\chi = \left( \ba{c} G^-_Y \\ G^{--}_\chi \\
 \frac{w+i G_{Z'}}{\sqrt{2}} \ea \right)
\eeq where $G_W^{\pm}$, $G_Z$, $G_Y^\pm$ and $G_{\chi}^{\pm\pm}$ are
the would-be Goldstone bosons  for the fields $W^\pm$, $Z$,
$Y^\pm$ and $\chi^{\pm\pm}$ respectively. The $\rho$ and $\eta$
components give origin to  the charged Higgs $H^+$, odd-$A^0$,
even-$H^0$ (all with  masses of the order of $w$, scale of energy
of the first symmetry breaking) and the light Higgs $h^0$ coming
from the electroweak scale.
 The Higgs fields $\rho^{++}$, $\eta_2^+$ and $Re(\chi^0)$ have a mass
proportional to the scale $w$.  The other fields of $\chi$ give
origin to the would-be Goldstone bosons of $\chi^{++}$, $Y^-$ and
$Z'$ . All the scalar fields, except for $h^0$, have masses of the order
of the first symmetry breaking $M_\chi$.

The covariant derivative can be written in terms of  the mass
eigenstates in the following way: \beqa D_\mu &=& \partial_\mu + i
e Q A_\mu
 + i \frac{g}{c_W} (T_{3L} - s_W^2 Q) Z_\mu +
i \frac{g}{\sqrt{1-3 t_W^2}} \left( \sqrt{3} t_W^2 (Q-T_3) +
T_8\right) Z'_\mu  \nonumber\\ & & + i \frac{g}{2} \left[ W^+_\mu
(T_1 - i T_2) + Y_\mu^- (T_4 - i T_5) + \chi_\mu^{--} (T_6 - i T_7) +
h.c.\right] \eeqa where the $T_i$ are the $SU(3)_L$ generators.
The gauge neutral fields are related in the following way: \beq
\left(\ba{c} W_\mu^3\\W_\mu^8\\B_\mu\ea\right)
 =
\left(\ba{ccc}
s_W & c_W & 0\\
- \sqrt{3} s_W & \sqrt{3} s_W t_W & \sqrt{1 - 3 t_W^2}\\
c_W \sqrt{1 - 3 t_W ^2} & -s_W \sqrt{1 - 3 t_W^2} & \sqrt{3} t_W
\ea\right) \left(\ba{c}A_\mu\\Z_\mu\\Z_\mu'\ea\right)\eeq
where $W_\mu^{3,8}$ and $B_\mu$
 are the gauge fields for the groups $SU(3)_L$
 and
$U(1)_X$ respectively. The $Z - Z'$ mixing is defined as
\beq
\left(\ba{c} Z_\mu\\Z'_\mu\ea\right)\left(\ba{cc}\cos\theta
 & \sin\theta\\
-\sin\theta & \cos\theta\ea\right) \left(\ba{c} Z_{1\mu}\\Z_{2\mu}\ea\right)
\eeq
Finally, the Yukawa lagrangian is: \beqa -{\cal{L}}_Y &=&
 \lambda_3 \,\bar{Q}_{3L}\, J_{3R}\, \chi + \lambda_{ij}\,
\bar{Q}_{iL} \,J_{jR} \,\chi^* + \lambda_{3a}\, \bar{Q}_{3L}
\,d_{aR}\, \rho +\lambda_{ia} \,\bar{Q}_{iL} \,u_{aR}
\,\rho^*\nonumber\\&& + \lambda'_{3a} \,\bar{Q}_{3L} \,u_{aR}
\,\eta+ \lambda_{ia} \,\bar{Q}_{iL}\, d_{aR} \,\eta^* + h_{mn}
\bar{f}_{mL} f^c_{nL}\eta^* + h.c.
  \eeqa
where the Yukawa constants $\lambda_{33}$ and $\lambda_3$ are
given by: \beq \lambda_{33} \sim  \frac{g}{\sqrt{2}}
\frac{m_b}{M_W} \;\;\;\;,\;\;\;\; \lambda_3 \sim \sqrt{2} g
\frac{m_{J_3}}{M_\chi}. \eeq

\section{331 Models and $Z \rightarrow b\bar{b}$}

Some of the Feynman rules that we need for the computation of $Z
\rightarrow b\bar{b}$ are shown in Figure 1. The new physics
diagrams that contribute to one loop order are shown in Figure 2.
We can write the one loop contribution to the width $\Gamma (Z
\rightarrow b\bar{b})$ as \cite{zbb} \beq \Delta\Gamma = i
\frac{e}{4 s_W c_W} \gamma_\mu (1-\gamma_5) Re(\delta^{SM}_b +
\delta^{NP}_b) \eeq with \beq \delta^{NP}_b = - \frac{\alpha}{4
\pi s_W} F_b \eeq and \beq F_b = \sum_j (I_j(x,y)-I_j(0,y)) =
I(x,y) -I(x,0). \eeq The variables $x$ and $ y$ take the value
$x=(m_f/M_G)^2$ and $y=(M_Z/M_G)^2$, $m_f$ is the mass of the
fermion circulating into the loop and $M_G$ is the mass of the
gauge field. The new physics contribution comes with $m_{J_3}$ and
$M_\chi$. The 331 model has a left-handed structure; this implies
that the one loop contribution of these fields has, for each
vertex,
 exactly the same  tensorial structure
as the SM contribution. Using Ref.\cite{zbb}, and substituting
$m_t \rightarrow m_{J_3}$, $M_W \rightarrow M_\chi$ and taking into
account the Feynman rules shown in Figure 1 we can compute the
$I_j(x,y)$ one loop functions:

\beqa I(x,y) &=& -\frac{5}{3} s_W^2 \left[ \frac{3}{2} +
\frac{2}{3} x + \left( 1-\frac{4}{3} x \right) f_{0a}+ \left(
\frac{1}{3} + y + \frac{1}{6} x (-4+x) \right) f_{2a}-
\right.\nonumber\\&&\left.\left( y+x +\frac{x^2}{2} \right) f_{1a}
\right] + s_W^2 \left[ -\frac{3}{4}c_W - 3 x + 2 \left(
 \frac{4}{3} c_W +x
 \right)
f_{0b}+
\left(
 x (1-x) + \right.\right.\nonumber\\&&\left.\left.\frac{c_W}{2}
 \left(
  y+\frac{1}{2} (1-x)
\right) \right) f_{2b} + x (c_W -1 +2 x) f_{1b} \right]
-\frac{3}{8} c_W + \frac{3}{2} x + \nonumber\\&&\left(
\frac{3}{4}c_W -x \right)
 f_{0b} +
 \left(
-\frac{1}{2} x (1-x) +\frac{c_W}{4} (y+\frac{1}{2}
 (1-x))
 \right)
 f_{2b}
+ \nonumber\\&& \left(
 \frac{1}{2} c_W-x
 \right)
  x f_{1b}+
\left(
1-\frac{2}{3} s_W^2
\right)
 x
 \left(
 -1+\frac{x}{1-x}+\frac{x^2 \ln x}{(1-x)^2}
 \right)
 +\nonumber\\&&
 \left(\frac{1}{2} + 2 s_W^2\right)
 \left(
 \frac{x}{1-x}+\frac{x^2 \ln x}{(1-x)^2}
 \right)
\eeqa where $x= (m_{J_3}/M_\chi)^2$ and $y=(M_Z/M_\chi)^2$. The
$f_{ia(b)}$ functions can be computed using Ref.\cite{zbb} and are
shown  in the Appendix.

The exotic
quarks $J_1, J_2$ do not couple
 to the bottom quark $b$. As we have already said, they do not mix with $J_3$ because
they have different electric charge. Besides, the gauge fields
$Z', Y$ do not contribute up to one loop. Scalar field
contributions can be summarized in the following way. All $\chi$
components are would-be Goldstone bosons, except for the third
(neutral) one that does not couples to the $b$-quark. The first
two components of $\eta$ and $\rho$ act exactly as two Higgs
doublets. The contributions due to $h^0, H^0, A^0$ are strongly
suppressed because the $b$-quark would be into the loop and,
moreover, the couplings would be proportional to $m_b$.

The $H^+$ contribution is suppressed with respect to the new
physics contributions coming with $m_{J_3}$ and $M_\chi$ because of
the suppression factors $(m_t/m_{H^+})^2$ and $(m_{J_3}/M_\chi)^2$
respectively, where the $H^+$-mass is of the order of $M_\chi$ but
$m_{J_3} \gg m_t$ in these models. The Yukawa term $\lambda_{33}
\bar{J_3} b\rho^{++}$ gives origin to the $\rho^{++}$
contribution, and it is suppressed because $\lambda_{33}\sim m_b$.
The $\eta_2^+$ does not  couples to particles in the loop.

From the  $Z - Z'$ mixing we obtain the relation \beq \delta
g_R^{Zbb} = - \frac{t_W s_W}{\sqrt{3 + 9 t_W^2}} \sin\theta \eeq

This formula implies that $\sin\theta \simeq - 0.3625$ is the
value that can explain the discrepancy for the asymmetry $\delta
g_R^{Zbb}$, as already mentioned in the introduction.

We will consider the following observables: $\Gamma(Z
\rightarrow b\bar{b})$, $R_b$, $R_c$, $R_l$ and $\Gamma_Z$. The new physics
contribution to them can be written as:
\begin{eqnarray}
\Gamma(Z\rightarrow
b\bar{b}) &=& \Gamma(Z\rightarrow b\bar{b})^{SM} (1+\delta^{NP}_b),\nonumber \\
\Gamma_Z &=& \Gamma_Z^{SM} [1+ Br (Z\rightarrow
b\bar{b})^{SM} \delta_b^{NP} ],\nonumber \\
R_b &=& R_b^{SM}[ 1 + \delta^{NP}_b
 (1-R_b^{SM})], \nonumber \\
R_c &=& R_c^{SM}[ 1 -  R_b^{SM} \delta^{NP}_b] ,\nonumber \\
R_l &=& R_l^{SM}[ 1 +  R_b^{SM} \delta^{NP}_b].
\end{eqnarray}
The theoretical and experimental values are given in Table 1
\cite{data}.
\begin{table}
\begin{center}
\begin{tabular}{||l||c|c||} \hline \hline
 & Experiment & Theory \\\hline \hline
 $\Gamma_Z[{\text GeV}] $ &  $2.4952 \pm 0.0023$ & $2.4972 \pm 0.0012$ \\ \hline
 $\Gamma (Z\rightarrow b\bar{b})[{\text MeV}]$ &
 $0.3775 \pm 0.0004 $ &$0.3758 \pm 0.0001 $\\\hline
$R_b$ & $0.21638 \pm 0.00066$ & $0.21544 \pm 0.00014$ \\ \hline
$R_c$ & $0.1720 \pm 0.0030$ & $0.17233 \pm 0.00005$ \\ \hline
$R_l$ & $20.784 \pm 0.043$ & $20.763 \pm 0.014$ \\\hline\hline
\end{tabular}
\end{center}
\caption{LEP observables directly sensitive to the $Zb\bar{b}$
vertex}\end{table} In order to determine the allowed region in the
space parameter for $M_{J_3}$ and $M_\chi$ we compute a $\chi^2$
fit with the parameters of Table 1 at $95\%$ confidence level. The
two  $A^b_{FB}$ and $A_b$ asymmetries are not taken into account
in our analysis because
 of the $3 \sigma$
discrepancy between the theoretical SM prediction  and the
measured values.

In Figures 3 and 4 we show the allowed 95\% CL regions delimited
by the two curves in the space parameter of $M_\chi$ and
$M_{J_3}$.

\section{Conclusions}

The sensitivity of precision  observables at the Z-pole to the
structure of the $Z\rightarrow b\bar{b}$ vertex allows to study
new physics effects. These can originate deviations from the
standard model prediction, and, comparing with the experimental
errors one can deduce some limits in the masses of bilepton and
exotic quarks in 331 models. New limits on these particles are
found in this way. We performed a naive analysis of the three triplets
scalar sector contribution of our model to the $Zb\bar{b}$-vertex
and we found that this contribution is small relative to the
one coming from diagrams with $\chi^{++}$ and $J_3$  in the loop.
For the 331 model we have an important equation for the bilepton
$\chi^{++}$ and the exotic quark $J_3$ masses from the Z-pole. This
implies that in the case of a bilepton  with a mass of the order of $700$ GeV
were found in future colliders then an exotic $J_3$ quark charged
$\pm 5/3$ and with a mass in the range  $1500 - 4000$ GeV should
also be found.


As this model has a left-handed structure it can not resolve the
$A_b$ and $A_{FB}^b$ asymmetries at the one loop level because the
contribution to the value of $g_L^{bbZ}$ is very small. On the
other hand, the mixing of the two neutral currents needs a mixing
angle of the order of $\sin\theta = -0.3625$ to explain the
asymmetry deviations, and this value is not compatible with the
value one can obtain from the analysis of the $Z$-pole observables
and the weak charge $Q_W$, which is of the order of $\sin\theta
\leq 10^{-3} - 10^{-4}$.

We acknowledge the financial support from CLAF,
COLCIENCIAS-Colombia and PEDECIBA-Uruguay.

\newpage

\section{Appendix}

The functions $f_{ia(b)} (x , y)$ we used to evaluate the
corrections, in the limit when $x=(m_f/M_G)^2 \rightarrow 0$ are:

\begin{eqnarray}
f_{0a} &=& - 2 + \ln(y) - i \pi \nonumber \\
f_{1a} &=& \frac{\ln^2(1+y)}{y} + \frac{2}{y}
Li_2(\frac{y}{1+y})-\frac{2}{y} \ln(y) \ln(1+y)+i \frac{ 2 \pi}{
y} \ln(1+y) \nonumber \\
f_{2a} &=& -\frac{4}{y} \left[-1+\ln(y)+\frac{1}{y}
\left(\frac{1}{2}
\ln^2(1+y)+Li_2(\frac{y}{1+y})-\ln(y)\ln(1+y)\right) \right]+ \nonumber \\
&& i \frac{ 4 \pi}{y}(1-\frac{1}{y}\ln(1+y)) \nonumber \\
f_{0b} &=& - 2 + \sqrt{1-4/y} \ln
\left(\frac{1+\sqrt{1-4/y}}{1-\sqrt{1-4/y}} \right)-i \pi
\sqrt{1-4/y} \nonumber \\
f_{1b} &=& \frac{4}{y} \left[Li_2 \left
(\frac{-2}{-1+\sqrt{1-4/y}}\right)+Li_2\left(\frac{2}{1+\sqrt{1-4/y}}\right)\right]
\nonumber \\
 f_{2b} &=& \frac{4}{y^2}\left[y+y \sqrt{1-4/y} \ln \left (
\frac{1-\sqrt{1-4/y}}{1+\sqrt{1-4/y}} \right)+2Li_2
\left(\frac{-2}{-1+\sqrt{1-4/y}} \right) \right. \nonumber \\
&&\left. +2Li_2 \left(\frac{2}{1+\sqrt{1-4/y}} \right) \right]+i 4
\pi \frac{\sqrt{1-4/y}}{y} \nonumber
\end{eqnarray}

\newpage

\begin{figure}
\includegraphics[scale=1.0]{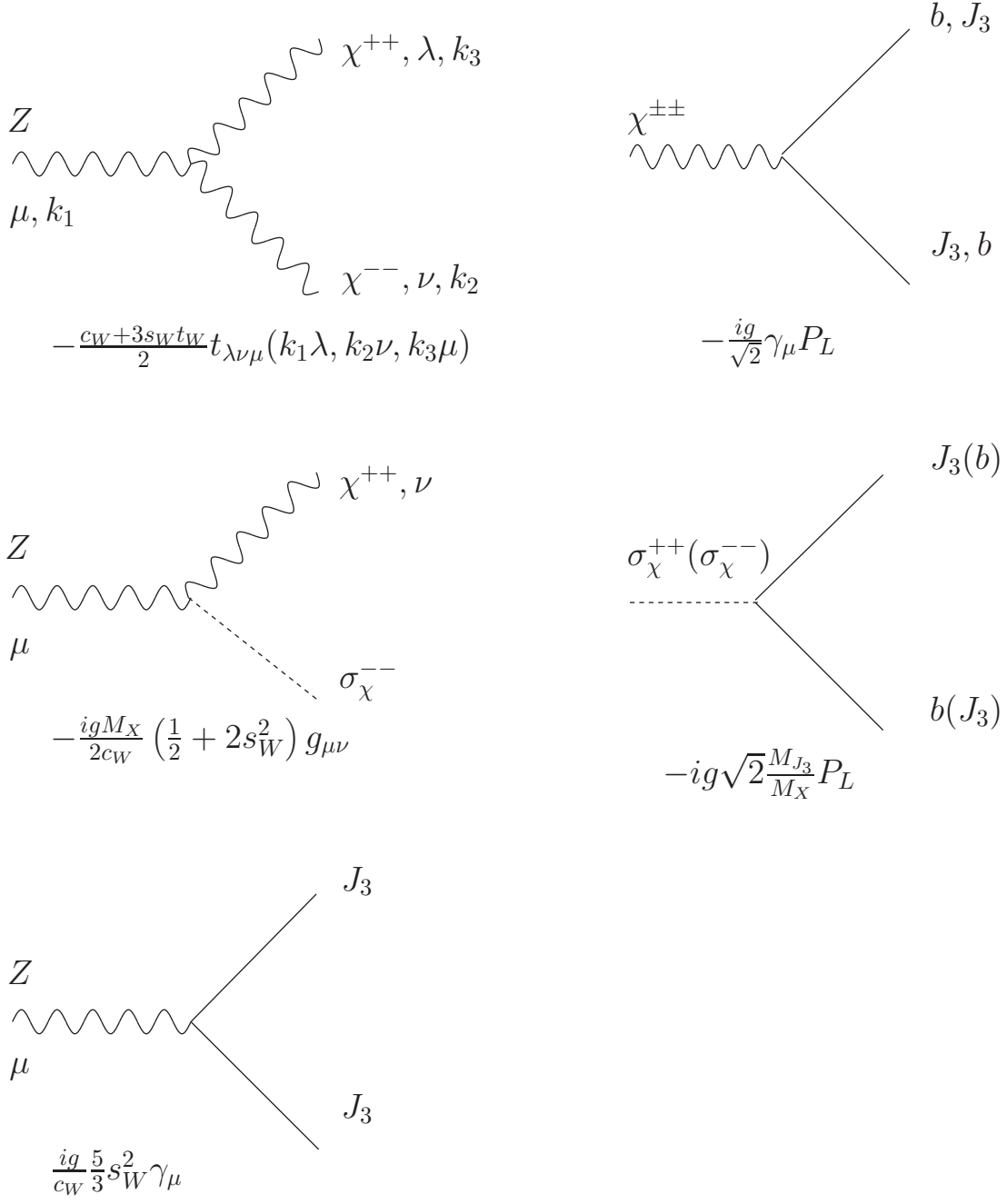}
\caption{Feynman rules for NP contributions to
$Z\rightarrow b\bar{b}$.}
\end{figure}

\newpage

\begin{figure}
\includegraphics[scale=1.0]{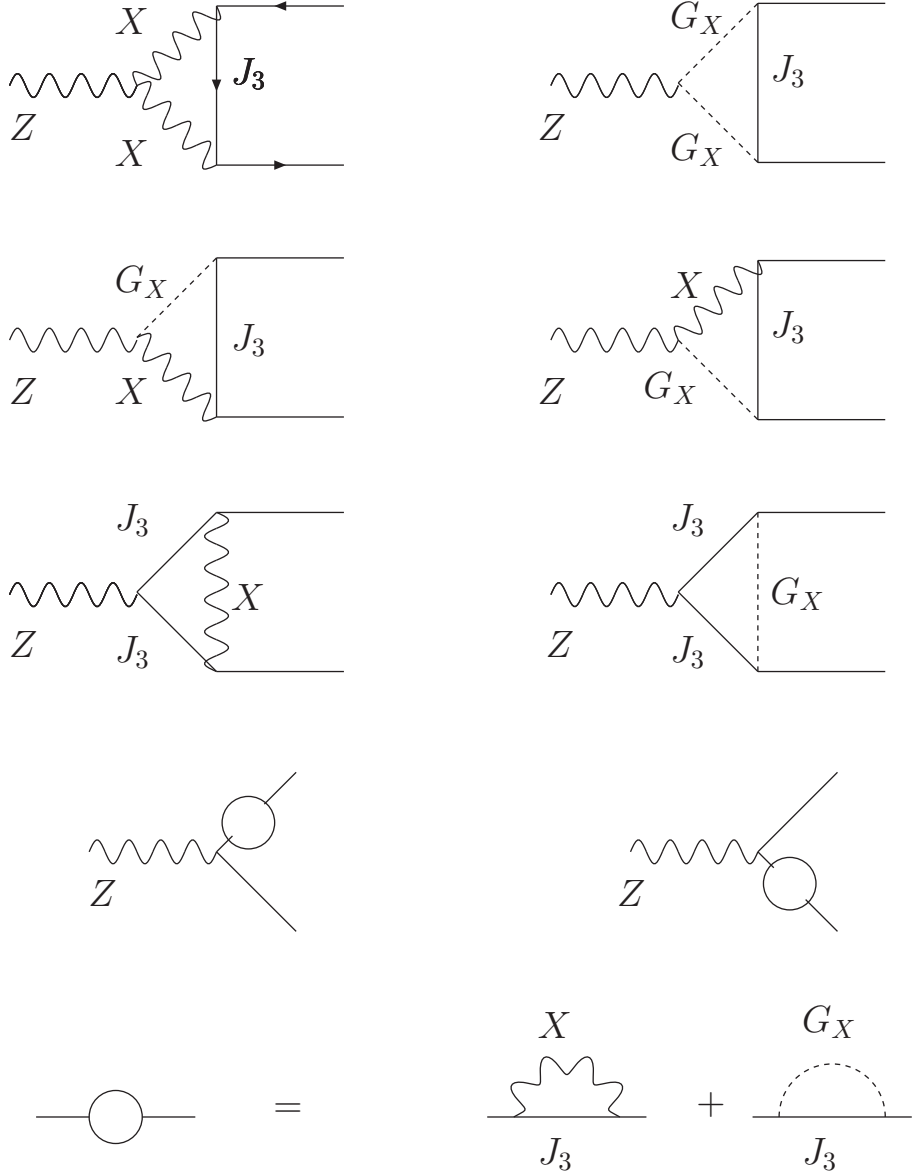}
\caption{One loop NP diagrams that contribute to
$Z\rightarrow b\bar{b}$. }
\end{figure}

\newpage

\begin{figure}
\includegraphics[scale=1.0]{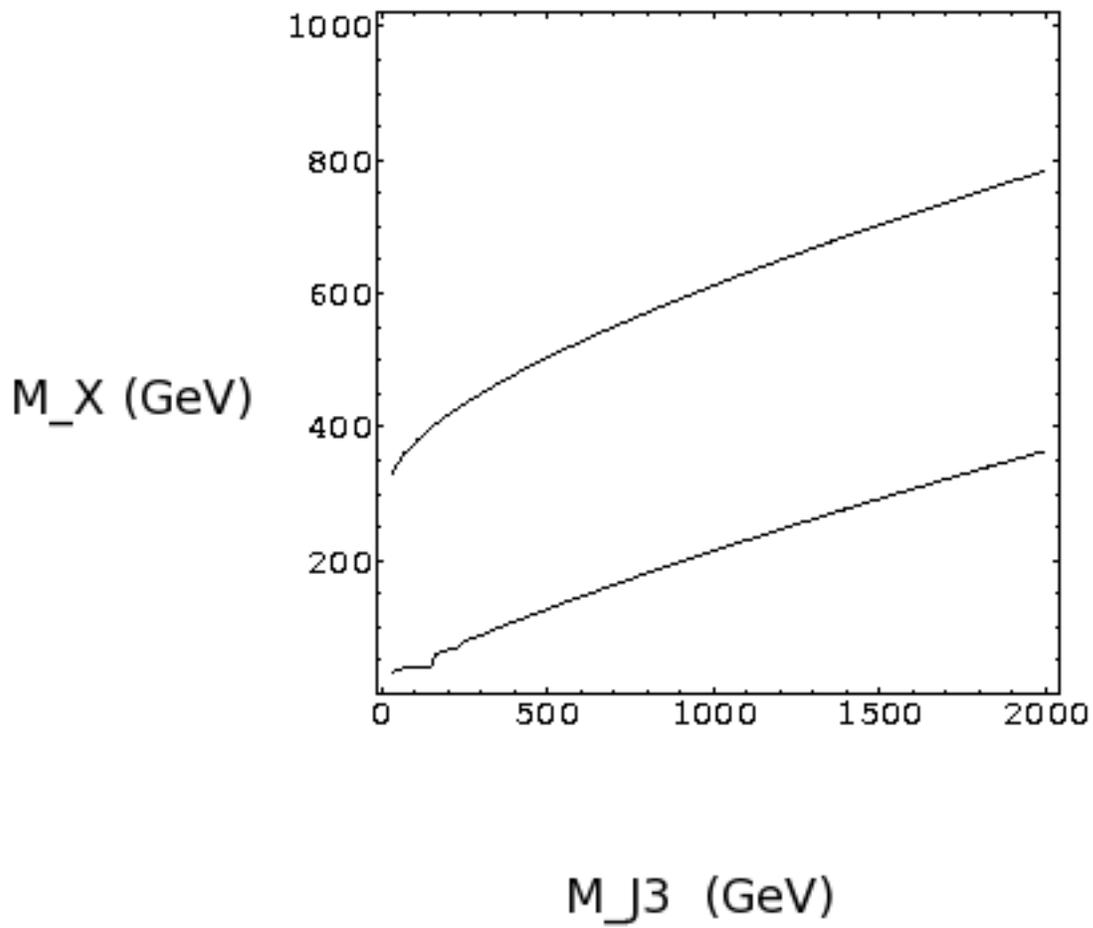}
\caption{Allowed region at 95\% CL for  $M_\chi, M_{J_3}$. }
\end{figure}

\begin{figure}
\includegraphics[scale=1.0]{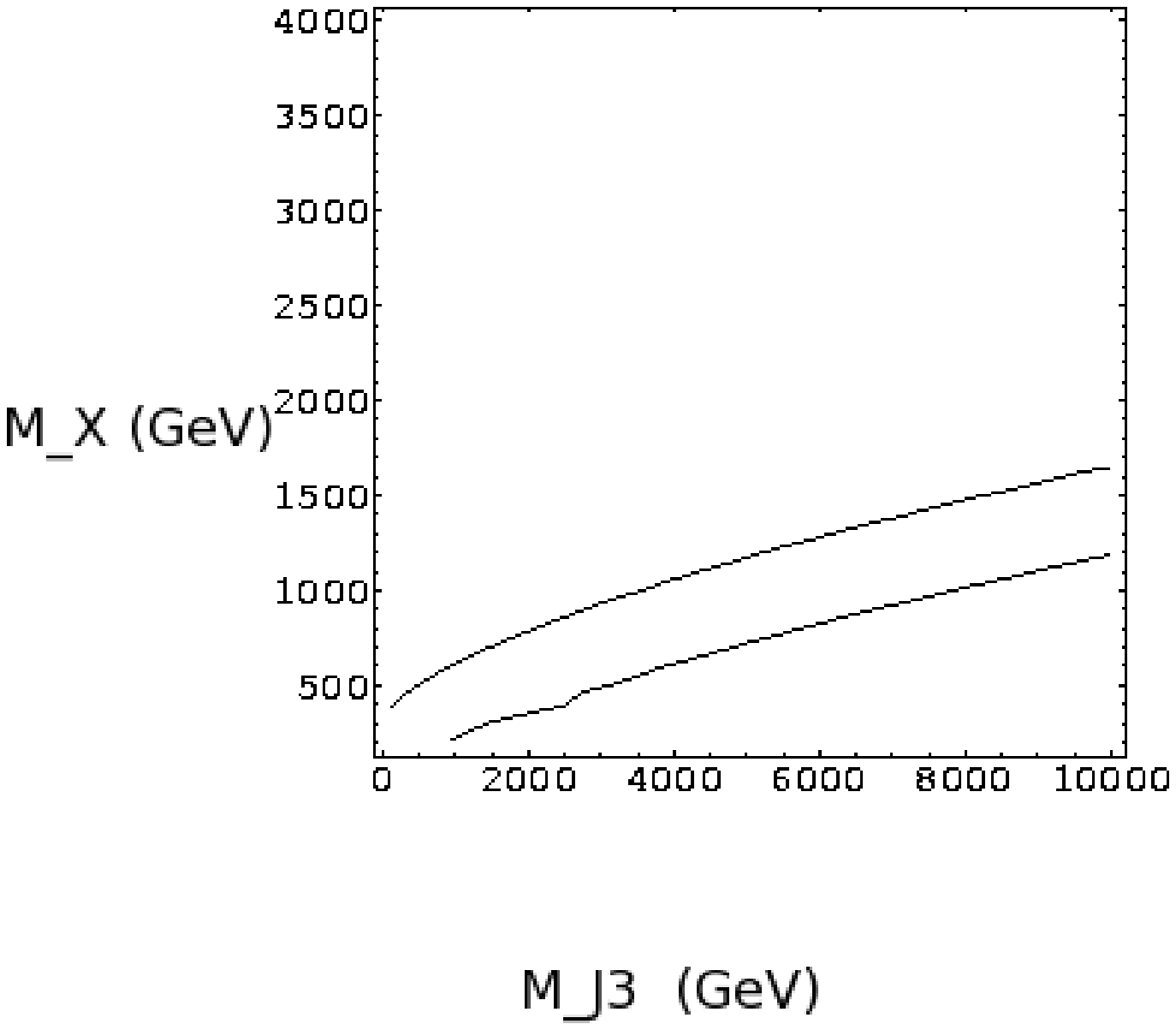}
\caption{Allowed region at 95\% CL for a wider region in the space
parameters $M_\chi, M_{J_3}$ plane. }
\end{figure}


\begin{thebibliography}{99}
%
\bibitem{sm}
S.L.Glashow, Nucl.\ Phys.\ {\bf 22}, 579 (1961); S. Weinberg,
 Phys.\ Rev.\ Lett. \ {\bf 19}, 1264 (1964); A. Salam, in {\it
 Elementary Particle Theory} (Nobel Symposium Nr. 8), edited by
N. Svartholm, Almqvist and Wiksell, Stckholm, Sweden (1968).
%
\bibitem{no}
Super Kamiokande Collab, Phys. \ Rev. \ Lett. \ {\bf 81}, 1562
(1998).
%
\bibitem{gim}
S.L. Glashow, J. Illiopulos and L. Maiani, Phys.\ Rev.\ D{\bf 2},
1285 (1970).
%
\bibitem{zbb}
J. Bernabeu, A. Pich and A. Santamaria,Nucl. \ Phys.\ B {\bf 363},
326 (1991).
%
\bibitem{top}
S. Abachi {\it et. al.}, Phys.\ Rev.\ Lett.\ {\bf 79}, 1203,
(1997); F. Abe {\it et al}, CDF Collab.,Phys.\ Rev.\ Lett.\ {\bf
80} ,1720 (1998).
%
\bibitem{higgs}G. Abbiendi {\it et al}
Phys.\ Lett.\  B {\bf 565}, 61 (2003).

%
\bibitem{preg}
P. Langacker, Phys.\ Rep.\ {\bf 72}, 185 (1981); H. Fritzch and Z.
Xing, Prog.\ Part.\ Nucl.\ Phys.\ {\bf 45}, 1 (2000).
\bibitem{higgslight}L.~Alvarez-Gaume, J.~Polchinski and M.~B.~Wise,
Nucl.\ Phys.\ B {\bf 221}, 495 (1983);H.P. Nilles, Phys.\ Rep.\
{\bf 110}, 1 (1984);H.E. Haber and G.L. Kane, Phys.\ Rep.\ {\bf
117}, 75 (1985).
%
\bibitem{lh}
N.~Arkani-Hamed, A.~G.~Cohen, E.~Katz and A.~E.~Nelson,
JHEP {\bf 0207}, 034 (2002);
N.~Arkani-Hamed, A.~G.~Cohen and H.~Georgi,
Phys.\ Lett.\ B {\bf 513}, 232 (2001);
N.~Arkani-Hamed, A.~G.~Cohen, E.~Katz, A.~E.~Nelson, T.~Gregoire
and J.~G.~Wacker,
JHEP {\bf 0208}, 021 (2002);
T.~Han, H.~E.~Logan, B.~McElrath and L.~T.~Wang,
Phys.\ Rev.\ D {\bf 67}, 095004 (2003).
%
\bibitem{randall}L. Randall and R. Sundrum, Phys.\ Rev.\ Lett.\ {\bf 83}, 3370 (1999);
A. Pomarol, Phys.\ Rev.\ Lett.\ {\bf 85}, 4004 (2000); A.
Arkani-Hamed, S. Dimopoulos and G. Dvali, Phys. \ Lett.\ B{\bf
429}, 263 (1998).
%
\bibitem{m331}
F. Pisano and V. Pleitez, Phys.\ Rev.\ D{\bf 46}, 410 (1992); P.H.
Frampton Phys.\ Rev.\ Lett.\ {\bf 69}, 2889 (1992).
%
\bibitem{flias}
R.A. D\' \i az, R. Mart\'\i nez and F. Ochoa, Phys.\ Rev.\ D {\bf
69}, 095009 (2004).
%
\bibitem{d1} S. Atag and K.O. Ozansoy, Phys.\ Rev.\ D{\bf 68}, 093008 (1993); Phys. \
 Rev.\ D {\bf 70}, 053001 (2004).
%
\bibitem{d2} P.H. Frampton and M. Harada, Phys.\ Rev.\ D {\bf 58}, 095013
(1998).
%
\bibitem{d3} L. Willmann {\it et al} Phys.\ Rev.\ Lett.\ {\bf 82}, 49 (1999).
%
\bibitem{d4} M.B. Tully and G.C. Joshi, Phys.\ Lett.\ B {\bf 466}, 333 (1993).
%
\bibitem{d5} D. London {\it et al},  Phys.\ Rev.\ D {\bf 59}, 075006 (1999).
%
\bibitem{exotic}
M.S. Chanowitz, Phys.\ Rev.\ Lett.\ {\bf 87}, 231802 (2001) and
Phys. \ Rev.\ D {\bf 60}, 073002 (2002).
%
\bibitem{dg}
G. Altarelli, F. Caravaglios, G.F. Giudice, P. Gambino and G.
Ridolfi, JHEP {\bf 06}, 018 (2001); P. Bomert, C.P: Burgess, J.M.
Cline, D. London and E. Nardi, Phys.\ Rev.\ D{\bf 54}, 4275
(1996); X. He and  G. Valencia, Phys.\ Rev.\ D {\bf 66}, 013004
(2002) and Phys.\ Rev.\ D{\bf 68}, 033011 (2003).
%
\bibitem{wch}
J. Erler and P. Langacker, Phys. \ Lett.\ B{\bf 456}, 68 (1999);
R. Mart\'\i nez, W. Ponce and L.A. S\'anchez, Phys.\ Rev.\ D 64,
075013 (2001) and Phys. \ Rev. \ D {\bf 65}, 055013 (2002).
%
\bibitem{foot}R. Foot, F. Hernandez,
 F. Pisano and V. Pleitez, Phys. Rev. {\bf D47}, 4158 (1993);
V. Pleitez and M.D. Tonasse, Phys. Rev. {\bf D48}, 2353 (1993);
T.V. Duong and E. Ma, Phys. Lett. {\bf B316}, 307 (1993).
%
\bibitem{long} H. N. Long, P. B. Pal, Mod. Phys. Lett. , {\bf A13}, 2355 (1998);
N. T. Anh, N. A. Ky, H. N. Long, Int. J. Mod. Phys., {\bf A16},
541 (2001).
%
\bibitem{data}S. Eidelman {\it et. al.}, Particle Data Group. Phys. Lett.
{\bf B592}, 1 (2004).






\end{thebibliography}
\end{document}